\documentclass[prb,reprint]{revtex4-1}
\usepackage[utf8]{inputenc}
\usepackage[english]{babel}
\usepackage{amsmath,amsfonts,amssymb}
\usepackage[T1]{fontenc}
\usepackage{url}
\usepackage[colorlinks=true, allcolors=blue]{hyperref}
\usepackage{graphicx,xcolor}
\usepackage{enumitem} 
\usepackage{tikz}
\usetikzlibrary{arrows.meta,decorations.pathmorphing,calc,patterns,datavisualization,datavisualization.formats.functions,angles,quotes}
\pgfdeclaredataformat{fresneldata}
    {}{}{#1   #2   #3   #4}
    {
    \pgfkeys{/data point/.cd, x=#1, y=#2, set=flat} \pgfdatapoint
    \pgfkeys{/data point/.cd, x=#1, y=#3, set=convex} \pgfdatapoint
    \pgfkeys{/data point/.cd, x=#1, y=#4, set=concave} \pgfdatapoint
    }{}{}
\pgfdeclaredataformat{fresneldata2}
    {}{}{#1   #2   #3   #4}
    {
    \pgfkeys{/data point/.cd, x=#1, y=#3, set=convex50} \pgfdatapoint
    \pgfkeys{/data point/.cd, x=#1, y=#4, set=concave50} \pgfdatapoint
    }{}{}

\begin{document}

\title{Analytical Fresnel laws for curved dielectric interfaces}

\author{Sebastian Luhn}
\author{Martina Hentschel}
\email[Corresponding author: ]{martina.hentschel@tu-ilmenau.de}
\affiliation{Institute for Physics, Technische Universit\"at Ilmenau, Weimarer Str. 25, 98693 Ilmenau}
\date{\today}

\begin{abstract}
Fresnel laws and the corresponding Fresnel reflection and transmission coefficients provide the
quantitative information of the amount of reflected and transmitted (refracted) light in dependence
on its angle of incidence. They are at the core of ray optics at planar interfaces. However, the
well-known Fresnel formulae do not hold at curved interfaces and deviations are appreciable when
the radius of curvature becomes of the order of several wavelengths of the incident light. This is of
particular importance for optical microcavities that play a significant role in many modern research
fields. Their convexly curved interfaces modify Fresnel's law in a characteristic manner. Most
notably, the onset of total internal reflection is shifted to angles larger than critical incidence \cite{Hentschel_Schomerus_PRE2002}. Here,
we derive analytical Fresnel formulae for the opposite type of interface curvature, namely concavely curved refractive index boundaries, that have not
been available so far. The accessibility of curvature-dependent Fresnel coefficients facilitates the analytical, ray-optics based description of light in complex mesoscopic optical structures that will be important in future nano- and microphotonic applications.
\end{abstract}

\maketitle
\section{Introduction}

The Fresnel equations, derived by Augustine-Jean Fresnel in 1823, quantify 
the amount of the reflected, $R$, and transmitted, $T=1-R$, intensity of a plane wave incident under a certain angle of incidence $\chi$ at a planar interface between two isotropic optical media of refractive indices $n_1$ and $n_2$ \cite{lipson_lipson_lipson_2010}. 
In their original form, they apply to flat interfaces with a relative index of refraction $n=n_1/n_2$. For internal reflection configurations ($n > 1$, reflection at the optically thinner medium) 
total internal reflection occurs above the critical angle of incidence given by $\chi_\mathrm{cr} = \arcsin 1/n$.

Whereas the effect of the curvature
is negligible in many cases when a ray optics description is adequate, it has to be taken into account when the radii $a$ of curvature of the interfaces become as small as several dozens or even several wavelengths $\lambda$ of the incident light to ensure a reliable description of the reflection and transmission process. This applies for example to optical microcavities with typical sizes of a few dozens micrometers across operated at infrared light \cite{Chang_microcavities}, \cite{Vahala_microcavities}. For example, the interface curvature affects, partly via the change in the Fresnel coefficients, the direction of the far field emission of the microcavities \citep{deformed-cylinders_Schwefel,KotikHentschel} as well as semiclassical corrections to the ray picture \citep{Hentschel_Schomerus_PRE2002,SchomerusHentschel_phase-space,nonHamiltonian,Unterhinninghofen_PRE2008,Unterhinninghofen_PRE2010,PS_EPL2014,PS_JOpt2017}. This implies in particular to deviations from Snell's law as a result of the so-called Fresnel-filtering effect \citep{TureciStone02}. These deviations have been studied in detail for convexly shaped interfaces based on the analytical Fresnel formulae available in this case \citep{Hentschel_Schomerus_PRE2002,annular_billiard}. 

\begin{figure}
 \begin{center}
    \includegraphics[width=0.28\textwidth]{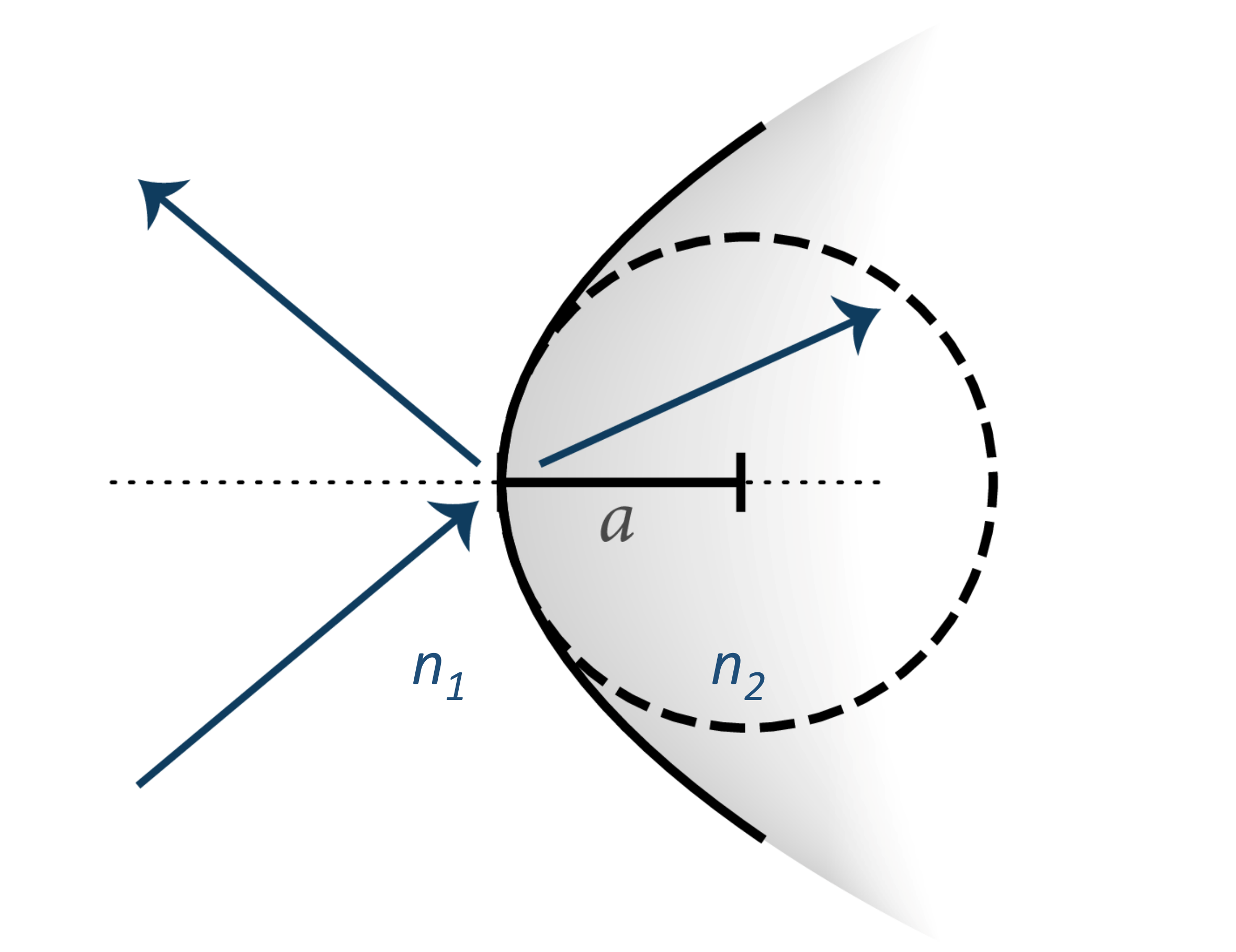}
 \end{center}
\vspace*{-0.8cm}
   \caption{Refraction of light at a concave optical interface between refractive indices $n_1$ and $n_2$ corresponding to a relative refractive index $n=n_1/n_2$. The curved interface can be \emph{locally} approximated by a cylinder with radius of curvature $a$.}
   \label{fig1}
\end{figure}

Here, we derive the missing analytical formulae for concavely curved boundaries as illustrated in Fig.~\ref{fig1}. They will provide the basis for a reliable ray-based description of photonic devices with convex or concave interfaces or complex boundaries that combine curved segments of both types. We point out that the analytical Fresnel laws that we present here apply to this general situation. To this end, the local radius of curvature $a$ of the device has to be used in the equations, i.e. the interface at the point of incidence of the incident ray is approximated by a cylinder (circle) of radius $a$, cf.~Fig. \ref{fig1}. 

The paper is organized as follows: We first state the analytical results for the Fresnel equations at curved interfaces (generalized Fresnel laws) both at convex and concave interfaces and discuss the deviations from the planar case. We then outline their derivation based on the transfer matrix approach that nicely illustrates the dual character of the convex-concave reflection situation.

\section{Fresnel reflection coefficients at curved interfaces}
\label{sec_results}
The Fresnel reflection coefficient for the reflected amplitude ratio $r$ of light propagating in a medium with relative refractive index $n > 1$ (internal reflection configuration) with an angle of incidence $\chi$ 
are known to read \cite{lipson_lipson_lipson_2010} for transverse magnetic (TM) and transverse electric 
(TE, note that we use the convention where TE  
features the Brewster angle at
$\chi_\mathrm{Br}=\arctan (1/n)$) polarization, respectively,
\begin{eqnarray}
r_p^\mathrm{TM} &=& \frac{ n \cos \chi -  \cos \eta}{n \cos \chi + \cos \eta} \\
r_p^\mathrm{TE} &=& - \frac{ n \cos \eta -  \cos \chi}{n \cos \eta + \cos \chi} 
\end{eqnarray}

Here, $\eta$ is the angle of the transmitted (refracted) light and given by 
Snell's law via $n \sin \chi = \sin \eta$.
\tikzdatavisualizationset {
	fresnel visualization/.style={
		scientific axes,
		visualize as line/.list={flat,convex,concave},
		style sheet = strong colors,
		style sheet = vary dashing,
		x axis = {label=$\sin\chi$, length=7.3cm},
		y axis = {label=Reflectance $R$,length=4cm},
		legend entry options/default label in legend path/.style= straight label in legend line, 
		legend = north west inside,
	},
	fresnel visualization ext/.style={
		fresnel visualization,
		visualize as line/.list={convex50,concave50},
	},
	legends1/.style= {
		flat={label in legend={text={planar $n=1.5$}}},
		convex={label in legend={text={convex/concave $n=1.5$}}},
		concave={label in legend={text={concave/convex $n=2/3$}}}
		
	},
	legends2/.style= {
		new legend entry={text={$n = 2/3$}},
		flat={label in legend={text={planar}}},
		convex={label in legend={text={convex/concave, $ka=15$}}},
		concave={label in legend={text={concave/convex $ka=15/n$}}} 
	},
	legends1 ext/.style= {
		new legend entry={text={$n = 1.5$}},
		flat={label in legend={text={planar}}},
		convex={label in legend={text={convex/concave, $ka=15$}}},
		concave={label in legend={text={concave/convex, $ka=15/n$}}},
		convex50={label in legend={text={convex/concave, $ka = 50$ }}},
		concave50={label in legend={text={concave/convex, $ka = 50/n$}}},
	},
}
\begin{figure}[htbp]
	\centering
	\begin{tikzpicture}[]
	\datavisualization[fresnel visualization ext,legends1 ext]
	data [read from file="TMAbsData.txt", format=fresneldata] 
	data [read from file="TMAbsData50.dat", format=fresneldata2];
	\end{tikzpicture}
	\vspace*{-0.3cm}
	\caption{Fresnel reflection coefficients $R = |r|^2, |r_\mathrm{cx}|^2, |r_\mathrm{cv}|^2$ for 
		TM-polarized light and $n=1.5$ at planar interface (black) and at curved interfaces with 
		$ka = 50$ (green), and $k a =15$ (red). Ray path reversal amounts to switching convex/concave and requires to renormalize the wavenumber $ka$ by $n$ (orange and blue curves).
		Note the delayed onset of total internal reflection as $ka$ is reduced and curvature effects become more dominant. 
	}
	\label{fig_reflTM}
\end{figure}
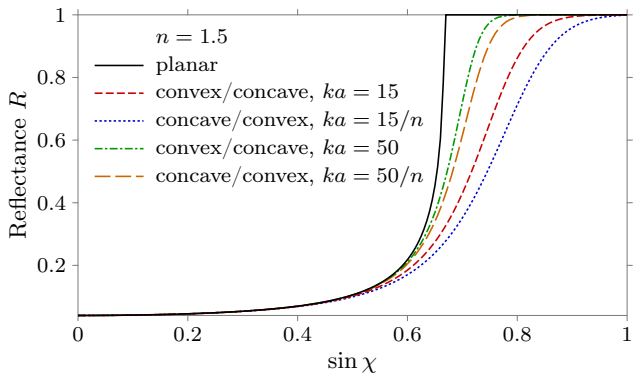
\begin{figure}[htbp]
	\centering
	\begin{tikzpicture}[]
	\datavisualization[fresnel visualization ext,legends1 ext]
	data [read from file="TEAbsData.txt", format=fresneldata]
	data [read from file="TEAbsData50.dat", format=fresneldata2];
	\end{tikzpicture}
	\vspace*{-0.3cm}
	\caption{Same as Fig.~\ref{fig_reflTM} above, but for TE-polarized light. 
		Note, however, that the Brewster angle reflectivity remains small but finite 
		at curved interfaces. As for the TM-case, deviations from the planar limit are important for reflection above the critical angle $\chi_\mathrm{cr} = \arcsin 1/n$ where curved interfaces are more leaky than planar boundaries. }
	\label{fig_reflTE}
\end{figure}
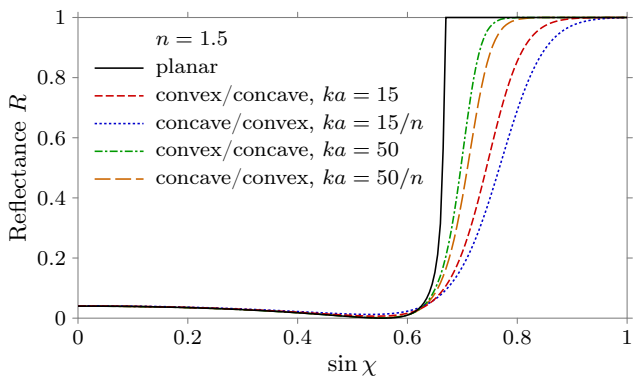

\subsection{Convex case}
We shall now see that the backbone structure of these equations is transferred to the curved interfaces. 
For the convex case, see \cite{Hentschel_Schomerus_PRE2002} and the alternative derivation below 
in Sec.~\ref{sec_transfermatrix}, we find the convex Fresnel reflection amplitude $r_\mathrm{cx}$ (assuming $n > 1 $) 
\begin{equation}
r_\mathrm{cx} = \frac{\cos \chi + i {\cal F}_m (k a)}{\cos \chi - i {\cal F}_m (k a)} \:.
\label{eq_refl_coeff_cx}
\end{equation}
Here, $ka$ is the dimensionless wave number given as product of the wave number $k$ in free space and the local radius of curvature $a$. 
The index $m$ is related to the angle of incidence via $m = nka \sin\chi=ka\sin\eta$\footnote{Note that $m$ is integer only for cylindrical resonators where it is the angular momentum quantum number ans is directly related to the order of the Hankel Functions\cite{annular_billiard}. In the general case we are interested here $m$ is related to the expectation value of the angle of incidence. Hankel functions of non integer order are well defined \cite{specialFunc}
and ensure the continuous behavior of the Fresnel coefficients, e.g. $J_\nu(x) = \protect\sum_{\nu=0}^{\infty}(-1)^\mu \frac{\left( x/2 \right)^{2\mu+\nu}}{\Gamma(\mu+1)\Gamma(\mu+\nu+1)}$}.
Furthermore, ${\cal F}_m$ is given for TM and TE polarization, respectively, as 
\begin{eqnarray}
  {\cal F}_m^\mathrm{TM}(z) &=& \frac{H_{m-1}^{1}(z)}{n \, H_m^{1}(z)} - \sin \chi \label{eq_calF_TM} \\
{\cal F}_m^\mathrm{TE}(z) &=& n^2 {\cal F}_m^\mathrm{TM}(z)\label{eq_calF_TE} \:.
\end{eqnarray}

\subsection{Concave case}
The result for the concave reflection amplitude $r_\mathrm{cv}$ is the central result of this paper, and
has a very similar, noteworthy structure,
\begin{equation}
r_\mathrm{cv} = \frac{\cos \chi - i {\cal F}_m^* (k a)}{\cos \chi + i {\cal F}_m^* (k a)} \label{eq_refl_coeff_cv}
\end{equation}
with the complex conjugation $()^*$ 
and ${\cal F}_m$ as given in Eqs.~(\ref{eq_calF_TM},\ref{eq_calF_TE}). It implies that the reflected intensity $R = |r^2|=r r^*$ is the same at a convex and concave interface, respectively. Note however, that reversal of the light path and the accompanying change from a concave to a convex interface boundary requires to renormalize the wavenumber by $n$, and the discussion of convex-concave duality below.

The results are illustrated in Figs.~\ref{fig_reflTM} and \ref{fig_reflTE} for TM and TE polarized light, respectively. The deviation from the planar case (black curve) is clearly visible and characterized by a much later onset of the regime of total internal reflection (i.e. for angles $\chi$ larger than the critical angle $\chi_{cr}$). This effect is more pronounced for smaller wavenumbers $ka$ (higher curvature). The planar case result is approached in the limit $ka \to \infty$. 

The reduced total internal reflection at curved boundaries implies a deterioration of the cavity quality ($Q$) factor and is thus important for many applications. 
We also point out that the drop of the reflectivity in TE polarization at the Brewster angle is less pronounced at all curved refractive index boundaries and does, in contrast to the planar case, not (quite) reach zero \citep{KotikHentschel,PS_JOpt2017}.

\subsection{External reflection configuration, $n < 1$}
Having considered the important case of optical microcavities where $n>1$, we now generalize the Fresnel equations for curved boundaries to relative refractive indices $n<1$ and find
\begin{eqnarray}
\label{eq_refl_coeff_thicker_cx}
{\tilde r}_\mathrm{cx} & = & - \frac{\cos \eta + i {\cal G}_m^* (k a)}{\cos \eta + i {\cal G}_m (k a)} \\
\label{eq_refl_coeff_thicker_cv}
{\tilde r}_\mathrm{cv} & = & - \frac{\cos \eta - i {\cal G}_m (k a)}{\cos \eta - i {\cal G}_m^* (k a)}
\end{eqnarray}
with
\begin{equation}
{\cal G}_m^\mathrm{TM}(z) = \frac{n \, H_{m-1}^{2}(z)}{H_m^{2}(z)} - \sin \eta 
 \:, \:\:
{\cal G}_m^\mathrm{TE}(z) = {\cal G}_m^\mathrm{TM}(z)\label{eq_calG_TMTE} / n^2 \:.
\end{equation}

The results are shown in Figs.~\ref{fig_reflTMLow} and \ref{fig_reflTELow}. As before, the reflectivity remains finite around the Brewster angle. The most striking difference to the planar case is the rather low reflectivity near grazing incidence at curved interfaces, confirming their larger leakage that we already observed for $n>1$. 

\subsection{Convex-concave duality}
The principle of ray-path reversal as well as the transfer matrix approach outlined below suggest to consider a convex interface 
together with its concave counterpart as two possible deviations from the planar case for a given $n$. 
This implies, however, a renormalization of the reference wavenumber $ka$ in the reversed situation by a factor $1/n$. These results are included in Figs.~\ref{fig_reflTM} -- \ref{fig_reflTELow}. In Figs.~\ref{fig_reflTM} and \ref{fig_reflTE} the renormalization increases the effect of curvature (the factor $n$ can be captured in a decrease from $a$ to $a/n$, or alternatively, in an increase from $\lambda$ to $n \lambda$) and the curves are further away from the ray limit $k a \to \infty$. For $n<1$, the same reasoning yields the opposite behavior, cf.~Figs.~\ref{fig_reflTMLow} and \ref{fig_reflTELow}.

To this end we point out 
a symmetry relation between the convex and concave reflectance {\it for a given order of the Hankel function $m$}: $|r_\mathrm{cx}(n\rightarrow n_2)|^2=|r_\mathrm{cv}(n\rightarrow n_2)|^2$, i.e., both coincide and deviate in the same manner from the planar case result. Note however that $r_\mathrm{cx}$ and $r_\mathrm{cv}$ differ in a phase 
such that $r_\mathrm{cx}(n\rightarrow n_2)=r_\mathrm{cv}^*(n\rightarrow n_2)$.
Deviations from this symmetry may occur when light {\it beams} that consist of numerous single rays are considered 
\citep{PS_EPL2014}.

\begin{figure}[htbp]
\centering
\begin{tikzpicture}[]
    \datavisualization[fresnel visualization,legends2]
    data [read from file="TMLowAbsData.txt", format=fresneldata];
\end{tikzpicture}
\vspace*{-0.3cm}
\caption{Same as Fig.~\ref{fig_reflTM} above, but now for the external reflection configuration, $n<1$. Note that the reflectivity at curved boundaries remains moderate even at grazing incidence. As before, the planar limit is reached as $ka$ is increased.  }
\label{fig_reflTMLow}
\end{figure}
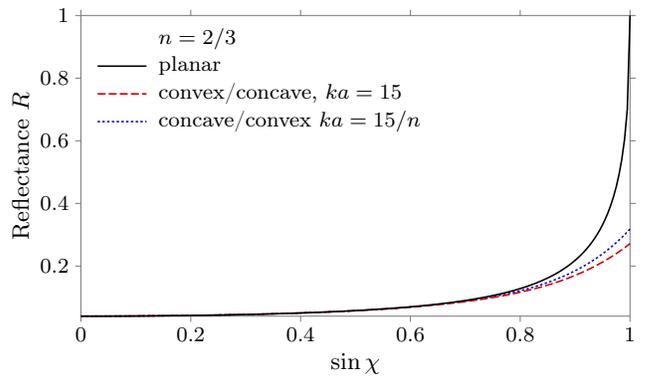

\begin{figure}[htbp]
\centering
\begin{tikzpicture}[]
    \datavisualization[fresnel visualization,legends2]
    data [read from file="TELowAbsData.txt", format=fresneldata];
\end{tikzpicture}
\vspace*{-0.3cm}
\caption{Same as Fig.~\ref{fig_reflTMLow}, but for TE-polarized light. 
}
\label{fig_reflTELow}
\end{figure}
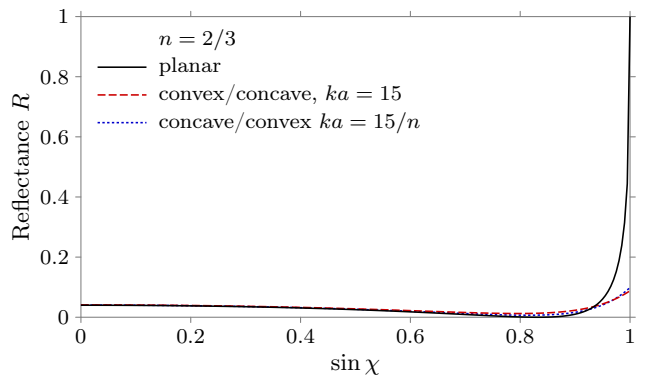

\section{Transfer matrices, resonances, and Fresnel coefficients}
\label{sec_transfermatrix}
\subsection{The transfer matrix}
\label{subsec_transfermatrix}
The transfer matrix relates incoming and outgoing wave amplitudes at (dielectric) interfaces \citep{born_wolf_1999}, cf.~Fig.~\ref{fig_transfermatrix}. Here $A_0$, $B_0$ and $A_1$, $B_1$ are the amplitudes of the incoming and outgoing waves, respectively, being related by the transfer matrix $M$ as 
\begin{align} \label{eq:transferMdef}
    \left(\begin{array}{c}
        A_1\\B_1
\end{array}\right) &=
    M\left(\begin{array}{c}
        A_0\\B_0
\end{array}\right) \;\textrm{where} & 
M&= \left(\begin{array}{cc}
        a & b \\
        b^* & a^*
    \end{array}\right)
\end{align}
Due to the presence of time-reversal symmetry, the $2\times2$ matrix $M$ takes the preceding form, where $a$ and $b$ are complex numbers. 
The relation to Fresnel coefficients is achieved when writing each outgoing amplitude in terms of a reflected and a transmitted contribution, see Fig.~\ref{fig_transfermatrix}, that are related by
\begin{eqnarray}
    A_1&=&r'B_1+tA_0 \\
    B_0&=&t'B_1+rA_0
\end{eqnarray}
Here $r$ and $t$ ($r'$ and $t'$) are the inner (outer) Fresnel reflection and transmission coefficients, respectively.
This yields the following 
representation of $M$ that holds independent of the curvature of the interface,
\begin{eqnarray} \label{eq:transferMfresnel}
    M= \left(\begin{array}{cc}
        1/t'^* & r'/t' \\
        r'^*/t'^* & 1/t' \:.
    \end{array}\right)
\end{eqnarray}

\tikzset{mediumstyle/.style={fill=black,opacity=0.13,draw=none}}
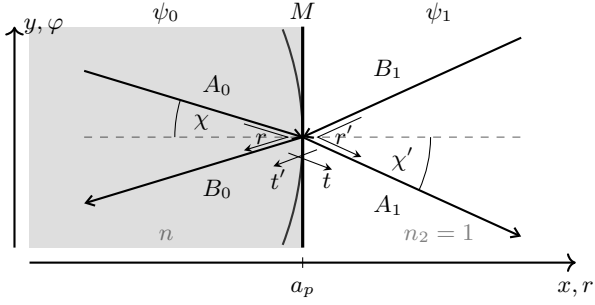
\begin{figure}[htbp]
\centering
    \def\iny{0.6}
    \def\outy{0.9}
    \def\halfw{0.8}
\begin{tikzpicture}[x=0.42\linewidth,y=0.17\linewidth,angle radius=1.70cm]
    \draw[dashed,draw=black!60] (-\halfw,0) coordinate (L) -- (\halfw,0) coordinate (R);
    \draw[mediumstyle] (-1,-1) rectangle (0,1);
    \def\arcAngle{21}
    \def\arcRadius{4cm}
    \draw[black!80,thick,radius=\arcRadius] (-\arcRadius,0) +(-\arcAngle:\arcRadius) arc [start angle=-\arcAngle, end angle=\arcAngle];
    \draw[very thick] (0,1) -- (0,-1);
        \coordinate (A0) at (-\halfw,\iny);
        \coordinate (A1) at (\halfw,-\outy);
        \coordinate (B0) at (-\halfw,-\iny);
        \coordinate (B1) at (\halfw,\outy);
        \coordinate (C) at (0,0);
        \begin{scope}[thick,black,arrows = -{Straight Barb[length=3pt,width=4pt]}]
        \draw (A0) -- node[auto]{$A_0$} (C);
        \draw (C)  -- node[auto]{$B_0$} (B0);
        \draw (B1) -- node[auto,swap]{$B_1$} (C);
        \draw (C)  -- node[auto,swap]{$A_1$} (A1);
    \end{scope}
    \begin{scope}[scale = 0.2,-stealth]
        \draw[xshift=-1cm] (-\halfw,\iny) -- (0,0) node[left=10pt,below=2.5pt,anchor=base] {$r$} -- (-\halfw,-\iny);
        \draw[xshift=+1cm] (+\halfw,\outy) -- (0,0) node[right=10pt,below=2.5pt,anchor=base] {$r'$} -- (+\halfw,-\outy);
    \begin{scope}[scale = 0.5,yshift=-3cm]
        \draw[xshift=+1cm] (-\halfw,\iny) -- (\halfw,-\outy) node[label=below:{$t$},left=2pt,above=2pt] {};
        \draw[xshift=-1cm] (+\halfw,\outy) -- (-\halfw,-\iny) node[label=below:{$t'$},right=2pt,above=2pt] {};
    \end{scope}
    \end{scope}
    \pic ["\small $\chi$", draw, angle eccentricity=0.8] {angle = A0--C--L};
    \pic ["\small $\chi'$", draw, angle eccentricity=0.8] {angle = A1--C--R};
    \begin{scope} [yshift=3pt]
    \node[anchor=base] at (-0.5,1) {$\psi_0$};
    \node[anchor=base] at ( 0.0,1) {$M$};
    \node[anchor=base] at (+0.5,1) {$\psi_1$};
    \node[anchor=base,opacity=0.5] at (-0.5,-1) {$n$};
    \node[anchor=base,opacity=0.5] at (+0.5,-1) {$n_2=1$};
\end{scope}
\begin{scope} [yshift=-0.2cm]
    \draw[thick,->] (-1,-1) -- (1,-1);
    \draw ([yshift=2pt] 0,-1) -- ([yshift=-2pt] 0,-1);
    \begin{scope} [yshift=-11pt]
        \node[anchor=base] at (1,-1) {$x,r$};
        \node[anchor=base] at (0,-1) {$a_{p}$};
    \end{scope}
\end{scope}
\draw[thick,->,xshift=-0.2cm] (-1,-1) -- (-1,1) node[anchor = west] {$y,\varphi$};
\end{tikzpicture}
\caption{Incident, transmitted, and reflected ray amplitudes are related via the transfer matrix at a curved interface (radius $a_p$) or a plane. Here, $n$ is the refractive index of the cavity that we assume to be embedded in air. See text for details. }
\label{fig_transfermatrix}
\end{figure}

\subsection{From the cavity transfer matrix to Fresnel coefficients}
In the following we will use 
the electromagnetic wave functions $\psi_0, \psi_1$ on either side of the interface (cf.~Fig.~\ref{fig_transfermatrix}) which are nothing else but the  
$z$-component of the electric (magnetic) field $E_z$ ($H_z$) in the TM (TE) case. 

Whereas in the planar case the $\psi$'s are plane waves and a plane divides space into two half planes \cite{lipson_lipson_lipson_2010}, a circle/cylinder takes this role in the presence of curvature. Consequently, we use Hankel functions as incoming and outgoing waves $\psi$ because they accommodate the cylindrical symmetry that we assume (locally) for a curved interface. 

The derivation of Fresnel coefficients at convex interfaces \citep{Hentschel_Schomerus_PRE2002} was based on finding the resonances in a disk cavity, and relating their real and imaginary part to the Fresnel reflection coefficient. 
Here, we use a conceptually different approach that is applicable to concave interfaces as well. 
To this end we extend the transfer matrix approach to a
Fabry-Perot-type situation with multiple reflections and mimic this behavior by including a perfect mirror, placed at $x, r = 0$. 
In the case of a planar interface, the plane-wave ansatz for 
$\psi_0$ and $\psi_1$ reads 
\begin{subequations} \label{eq:planeResAnsatz}
\begin{eqnarray}
    \psi_0(x) &=& \frac{I}{2}\left( e^{in_xk_xx} + e^{-in_xk_xx} \right) \nonumber \\
    		  &= & I\cos (n_xk_xx) \label{eq:planeInner} \\
    \psi_1(x) &=& e^{-ik_xx} + S e^{ik_xx} \label{eq:planeOuter} \:.
\end{eqnarray}
\end{subequations}
The factor $I$ describes (twice) the wave amplitude inside 
the resonator and $S$ is the transmitted (scattered) amplitude leaving the system. 
The effective refractive index $n_x$ is given as $n_x = \frac{\tan \chi'}{\tan \chi}$ where $\chi$ and $\chi'$ are defined in Fig.~\ref{fig_transfermatrix}.

Following the procedure outlined in Sec.~\ref{subsec_transfermatrix} above, the new transfer matrix $M'$ at the position $x=a_{p}$ reads
\begin{eqnarray}
    M'&=& \frac{1}{t'}
    \left(\begin{matrix}
        e^{i(n_x-1)k_xa_{p}} & r'e^{-i(n_x+1)k_xa_{p}} \\
        r'e^{i(n_x+1)k_xa_{p}} & e^{-i(n_x-1)k_xa_{p}}
    \end{matrix}\right) \:.
        \label{eq_17}
\end{eqnarray}
The analogy to the single-transmission case is established by introducing resonance-dressed Fresnel coefficients $r_r, r'_r$ and $t_r, t'_r$, characterized by the presence of an additional phase, namely
\begin{align} 
    r_r &= re^{2in_xk_xa_{p}} & r_r' &= r'e^{-2ik_xa_{p}} 
    \label{eqn17}\\
    t_r &= te^{i (n_x-1)k_xa_{p}} & t_r' &= t'e^{i(n_x-1)k_xa_{p}} 
    \label{eqn18}
\end{align}
A straightforward explanation of the resonance formation is gained when expressing the amplitude $I$ in terms of a geometric series of resonance-dressed Fresnel coefficients, nicely illustrating the successive reflections of light rays, 
\begin{equation} \label{eq:I2fres}
    \frac{I}{2} = \frac{t'_r}{1-r_r} = t'_r+t'_rr_r+t'_rr_r^2+t'_rr_r^3+\cdots \:.
\end{equation}
Note that the wave numbers $k$ at resonance can be obtained by solving the equation $1-r_r = 0$ and that $|I/2|^2$ will oscillate with its maxima reached at resonant wave numbers $k$.

\subsection{Fresnel coefficients at curved interfaces}
We adapt the ansatz for the planar cavity (Eqs. (\ref{eq:planeResAnsatz}a,b)) to the rotationally invariant case relevant for 
curved interfaces:
\begin{subequations} \label{eq:psibasesolution}
\begin{align}
    \psi_0(r,\varphi) &= \frac{I}{2} \left( H_m^{1}(nkr) + H_m^{2}(nkr) \right) =I J_m(nkr) e^{im\varphi} \\
    \psi_1(r,\varphi) &= (H_m^{2}(kr) + SH_m^{1}(kr))e^{im\varphi} \:.
\end{align}
\end{subequations}
Note that we switch from this convex case to the concave situation by exchanging incoming and outgoing Hankel functions (i.e., indices 1 and 2) while properly accounting for the scaling factor $n$, whereas 
the distinction between TM and TE polarized light originates in the well-known difference in the transition conditions \cite{lipson_lipson_lipson_2010}:
\begin{subequations}
    \begin{eqnarray}
    \psi(a_p-0) &=& \psi(a_p+0) \label{eq:stetig}\\
    \psi'(a_p-0) &=& \begin{cases}{}
        \psi'(a_p+0) &\text{TM} \\
        n^2\psi'(a_p+0) &\text{TE} \: .
    \end{cases} \label{eq:diffbar}
\end{eqnarray}
\end{subequations}

The transfer matrix takes then the following form
\begin{align} \label{eq:transferMatrixCurved}
    M^{(TM)} = &\frac{1}{D} \left( \begin{matrix}
            D_{12} & D_{22} \\
            -D_{11} & -D_{21}
    \end{matrix} \right) \\ 
    \begin{split} \label{eq:tetransferMatrixCurved}
        M^{(TE)} = & M^{(TM)} 
     + \frac{\sin \chi}{D} \left( n - \frac{1}{n} \right) \left( \begin{matrix}
            -Q_{12} & -Q_{22} \\
            Q_{11} & Q_{21}
    \end{matrix} \right)
    \end{split}
\end{align}

with (realizing that $a_p$ takes the role of $a$, and $\{\alpha,\beta\}$ $\in$ $\{1,2\}$)
\begin{eqnarray}
    D_{\alpha \beta} &=& H^{\alpha}_m(nka)H^{\beta}_{m-1}(ka)-nH_{m-1}^{\alpha}(nka)H_m^{\beta}(ka) \label{eq:Dab} \\ 
    D &=&  4 / (i \pi ka) \\ 
    Q_{\alpha \beta} &=& H_m^{\alpha}(nka) H_m^{\beta}(ka) \label{eq:Qab}
\end{eqnarray}

By analyzing the resonance-dressed (or multiple) reflection coefficients obtained from 
	Eqs.~(\ref{eq:transferMatrixCurved}, \ref{eq:tetransferMatrixCurved}), we find the following additional phases in the Fresnel reflection coefficients (cf.~Eqs.~(\ref{eqn17},\ref{eqn18}) for the planar case):
\begin{subequations}
    \begin{align}
        \text{convex:}\:\:\: \frac{H^1_m(nka)}{H^2_m(nka)} \:, \qquad
        \text{concave:}\:\:\: \frac{H^2_m(ka)}{H^1_m(ka)}
    \end{align}
\end{subequations}

The resulting reflection coefficient at a {\bf convex} interface reads thus 
\begin{equation} \label{eq:konvexReflecionCoeff}
    r^{(TM)} = -\frac{\frac{H_{m-1}^{1}(ka)}{H_m^{1}(ka)} - n \frac{H_{m-1}^{1}(nka)}{H_m^{1}(nka)}}{\frac{H_{m-1}^{1}(ka)}{H_m^{1}(ka)} - n \frac{H_{m-1}^{2}(nka)}{H_m^{2}(nka)}}
\end{equation}
This coefficient will not oscillate when changing the argument $nka$. It is therefore suitable to describe open segment boundaries, in consistency with allowing non-integer $m = nka \, sin \chi$ (the integer-$m$ constraint applies formally only in the presence of rotational symmetry). 

We proceed by simplifying the expressions using the large-argument approximation
\begin{equation}
\frac{H_{m-1}^{1}(nka)}{H_m^{1}(nka)} \approx e^{- i \left( \chi - \frac{\pi}{2} \right)} = \sin\chi + i\cos\chi \:,
\label{ratioH}
\end{equation}
where $\chi= {\rm arcsin} (m/nka)$ as before. 
This relation can be derived using a somewhat less-common approximation for the Hankel function \cite{transFunc} that holds when $m$ and the argument $nka$ of the Hankel function are of similar order as in the present case, namely  
\begin{equation}
    H^1_m(z) = \frac{\sqrt{2/\pi}}{\sqrt{z\cos\chi}}e^{i\varphi_m(z)}\left( 1- \frac{b_1(z)}{z\cos\chi} + O(z^{-2}) \right)
\end{equation}
where
\begin{equation}
    \varphi_m(z)=z\cos\chi-m \left( \frac{\pi}{2}-\chi \right)-\frac{\pi}{4}
\end{equation}
and
\begin{equation}
    b_1(z)=\frac{1}{8}+\frac{5}{24}\tan^2\chi \:.
\end{equation}
Equation (\ref{ratioH}) follows in the limit of large $m$ and $nka$, and eventually we find indeed the results stated in Eqs. (\ref{eq_refl_coeff_cx}) and 
 (\ref{eq_refl_coeff_thicker_cx}). 

In analogy, we find for the {\bf concave} case 
\begin{equation} \label{eq:konkavReflecionCoeff}
    r'^{(TM)} = -\frac{\frac{H_{m-1}^-(ka)}{H_m^-(ka)} - n \frac{H_{m-1}^-(nka)}{H_m^-(nka)}}{\frac{H_{m-1}^+(ka)}{H_m^+(ka)} - n \frac{H_{m-1}^-(nka)}{H_m^-(nka)}}\:.
\end{equation}
We proceed by simplifying the expressions using the previously introduced approximation for the argument $ka$ being in the order of $m$, and  $\eta= {\rm arcsin} (m/ka)$, 
\begin{subequations}
    \begin{align} \label{eq:approximationl}
        \frac{H_{m-1}^{1}(ka)}{H_m^{1}(ka)} &\approx e^{- i \left( \eta - \frac{\pi}{2} \right)} = \sin\eta + i\cos\eta \:, \\
        \frac{H_{m-1}^{2}(ka)}{H_m^{2}(ka)} &\approx e^{+ i \left( \eta - \frac{\pi}{2} \right)} = \sin\eta - i\cos\eta \:.
    \end{align}
\end{subequations}
From this, Eqs.~(\ref{eq_refl_coeff_cv}) and (\ref{eq_refl_coeff_thicker_cv}) follow.
For the derivation of the TE-case results in Chapter \ref{sec_results}, we proceed as
before but use the appropriate transfer matrix Eq. \eqref{eq:tetransferMatrixCurved}.

To {\bf summarize}, we have completed the picture of Fresnel coefficients at generic curved interfaces by deriving a formulae for the concave case in addition to the previously known convex case coefficients, and by illustrating the power of the transfer matrix approach to this problem.

\section{Funding Information}

M.H. acknowledges funding in the DFG Emmy Noether program. 
We thank Juluis Kullig, Jakob Kreismann, Henning Schomerus, and Stefan Sinzinger for stimulating discussions.
\bibliographystyle{unsrtnat}
\bibliography{optics}
\end{document}